\documentclass[preprint,letterpaper,showpacs]{revtex4}
\usepackage[T1]{fontenc}
\usepackage{amsfonts,amsmath,amssymb}

\def\bra#1{\mathinner{\langle{#1}|}}
\def\ket#1{\mathinner{|{#1}\rangle}}

{\catcode`\|=\active 
  \gdef\Braket#1{\left<\mathcode`\|"8000\let|\BraVert {#1}\right>}}
\def\BraVert{\egroup\,\mid@vertical\,\bgroup}
{\catcode`\|=\active
  \gdef\set#1{\mathinner{\lbrace\,{\mathcode`\|"8000\let|\midvert #1}\,\rbrace}}
  \gdef\Set#1{\left\{\:{\mathcode`\|"8000\let|\SetVert #1}\:\right\}}}
\def\midvert{\egroup\mid\bgroup}
\def\SetVert{\egroup\;\mid@vertical\;\bgroup}

\newcommand{\R}{\mathbb{R}}
\newcommand{\Z}{\mathbb{Z}}
\newcommand{\C}{\mathbb{C}}

\newcommand{\ad}{\mathrm{ad}}
\newcommand{\comm}[2]{\left[ #1 , #2 \right]}
\newcommand{\calA}{\mathcal{A}}
\newcommand{\calH}{\mathcal{H}}
\newcommand{\calC}{\mathcal{C}}

\begin{document}

\title{\bf Deformed Kac-Moody and Virasoro Algebras}

\author{A. P. Balachandran}
\email{bal@phy.syr.edu}

\affiliation{Department of Physics, Syracuse University \\
Syracuse, NY, 13244-1130, USA}

\author{A. R. Queiroz}
\email{amilcar@fma.if.usp.br}

\affiliation{Centro Internacional de F\'{\i}sica da Mat\'{e}ria Condensada,
 Universidade de Bras\'{\i}lia, C.P. 04667, Bras\'{\i}lia, DF, Brazil}

\author{A. M. Marques}
\email{amarques@fma.if.usp.br}

\author{P. Teotonio-Sobrinho}
\email{teotonio@fma.if.usp.br}

\affiliation{Instituto de F\'{\i}sica, Universidade de S\~{a}o Paulo \\
C.P. 66318, S\~{a}o Paulo, SP, 05315-970, Brazil}

\preprint{SU-4252-837}

\begin{abstract}
  Whenever the group $\R^n$ acts on an algebra $\calA$, there is a method
  to twist $\cal A$ to a new algebra $\calA_\theta$ which depends on an
  antisymmetric matrix $\theta$ ($\theta^{\mu \nu}=-\theta^{\nu
    \mu}=\mathrm{constant}$). The Groenewold-Moyal plane
  $\calA_\theta(\R^{d+1})$ is an example of such a twisted algebra. We
  give a general construction to realise this twist in terms of $\calA$
  itself and certain ``charge'' operators $Q_\mu$. For
  $\calA_\theta(\R^{d+1})$, $Q_\mu$ are translation generators. This
  construction is then applied to twist the oscillators realising the
  Kac-Moody (KM) algebra as well as the KM currents. They give different
  deformations of the KM algebra. From one of the deformations of the KM
  algebra, we construct, via the Sugawara construction, the Virasoro
  algebra. These deformations have implication for statistics as well.
\end{abstract}

\maketitle

\section{Introduction}

\subsection{\it  Preliminaries}

Let $(M,g)$ be a Riemannian manifold with metric $g$. Suppose that
$\R^N\;(N\geq 2)$ acts as a group of isometries on $M$. Then $\R^N$ acts
on the Hilbert space $L^2(M,d\mu_g)$ of square integrable functions on
$M$. The volume form $d\mu_g$ for the scalar product on $L^2(M,d\mu_g)$
is induced from $g$. Hence this action of $\R^N$ is unitary. 

We are interested just in the unitarity of the $\R^N$ action. Hence we can
weaken the isometry condition. That is because the action of $\R^N$
remains unitary if it leaves only $d\mu_g$ invariant without necessarily
leaving the metric $g$ itself invariant.

Let $\lambda = (\lambda_1,\ldots,\lambda_N)$ label the unitary irreducible
representations (UIR's) of $\R^N$. Then we can write
\begin{equation}\label{eq:1}
L^2(M,d\mu_g) = \bigoplus_\lambda \calH^{(\lambda)} \;,
\end{equation}
where $\R^N$ acts by the UIR $\lambda$, or in case of multiplicity, by
direct sums of the UIR $\lambda$, on $\calH^{(\lambda)}$. $\R^N$ being
noncompact, we may have to write the direct sum as the direct
integral. But, as this issue is not important here, we will use the
summation notation. 

\subsection{\it The Twist}

There is a general way of twisting the algebra of 
functions on $M$
using the preceding structure.We now explain it.

If $a=(a_1,\ldots,a_N) \in \mathbb{R}^N$ and the group law is 
addition,we choose $\lambda$ such that $\lambda:a \rightarrow 
e^{i\lambda \cdot {a}}$. Let
 $f_\lambda$ and $f_{\lambda'}$ be two
smooth functions in $\calH^{(\lambda)}$ and $\calH^{(\lambda')}$, Then
under the pointwise multiplication
\[ f_\lambda \otimes  f_{\lambda'} \rightarrow f_{\lambda}\,f_{\lambda'}\;,
\]
where 
\[ (f_{\lambda}\,f_{\lambda'})(p) = f_{\lambda}(p)\,f_{\lambda'}(p). \]
With $p$ a point of $M$, we have that 
\begin{equation}
  \label{eq:2}
f_{\lambda}\,f_{\lambda'} \in \calH^{(\lambda + \lambda')}\;.  
\end{equation}
Now suppose that $\theta$ ($\theta^{\mu \nu} = - \theta^{\nu \mu} =
\mathrm{constant}$) is an antisymmetric constant matrix in the space of
UIR's of $\R^N$. We can use it to twist the pointwise product to 
the
$*$-product $*_\theta$ depending on $\theta$ where
\begin{equation}
  \label{eq:3}
   f_\lambda *_\theta f_{\lambda'} =  f_\lambda \;  f_{\lambda'} \;
   e^{-\frac{i}{2} \lambda_\mu \theta^{\mu \nu} \lambda'_\nu}\;.
\end{equation}
The resultant algebra $\calA_\theta(M)$ is associative because of (\ref{eq:2}).

This algebra has been reviewed and developed by Rieffel \cite{RIEFFEL} and
many others. $\calA_\theta(\R^d)$, the Moyal plane, is a special case of this algebra. In
recent times, Connes and Landi \cite{CONNESLANDI} and Connes and Dubois-Violette
\cite{CONNESDV}, have constructed the full noncommutative geometry for
special cases of this algebra.

The discussion above  shows that what is pertinent for the twist is not
the commutative nature of the underlying algebra. Rather, it is
sufficient to have an associative algebra $\calA$ graded by 
$\lambda = (\lambda_1, \ldots, \lambda_N) \in \R^N \; (N \geq 2)$:
\begin{equation}
  \label{eq:grade1}
 \calA = \bigoplus \calA^{(\lambda)}\; ,\hspace{10mm}
\calA^{(\lambda)}\calA^{(\lambda')} \in \calA^{(\lambda + \lambda')}\;.
\end{equation}
Then we can twist it to $\calA_\theta$:
\[ \alpha_\lambda *_\theta \alpha_{\lambda'}  = \alpha_\lambda \;
\alpha_{\lambda'}\; e^{-\frac{i}{2}\lambda_\mu \theta^{\mu \nu}
  \lambda'_\nu}\;, \hspace{10mm} \alpha_{\lambda,\lambda'} \in \calA^{(\lambda),(\lambda')}\;.
\]
We will illustrate this deformation as well using bosonic creation-annihilation
operators and the Kac-Moody (KM) algebra\footnote{In the mathematical literature on KM
  and Virasoro algebras, their twists are understood differently. (We thank
  Prof. V. Dobrev for pointing this out to us). Hence for these algebras,
  we replace the phrase ``twist'' by the phrase ``deformation''.}.

It is also possible to twist an associative algebra $\calA$ to an
associative algebra $\calA_\theta$ on which $\mathbb{R}^N$ acts even if
this action is not unitary. If $V_\mu$, with $\mu=1,...,N$ is a basis of
vector fields for the Lie algebra of $\mathbb{R}^N$, and
$\alpha,\beta\in\calA$, then the twisted product is the $*$-product 
\begin{equation}
  \label{eq:star-product1}
  \alpha *_\theta \beta=\alpha e^{\overleftarrow{V}_\mu \theta^{\mu \nu} \overrightarrow{V}_\nu} \beta.
\end{equation}
However, if the $\mathbb{R}^N$ action is not unitary, we may encounter
problems in physical applications.

We remark that the grading lattice in (\ref{eq:grade1}) can even be periodic. Thus suppose that 
\[
\calA^{(\lambda_1,\lambda_2,\ldots,\lambda_l+M_l,\lambda_{l+1},\ldots,\lambda_N)}
= \calA^{(\lambda_1,\lambda_2,\ldots,\lambda_N)} \] where $M_l$ is the
period in direction $l$. Then it is enough for consistency that twisting
matrix $\theta$ in (\ref{eq:3}) satisfies
\[ e^{-\frac{i}{2} M_l \theta^{l \nu} {\lambda'}_\nu} = 1 \hspace{10mm}(\textrm{no $l$
  sum})\]
which implies 
\[ \theta^{l\nu} {\lambda'}_\nu = \frac{4\pi}{M_l}\;N\;, \hspace{10mm}N \in
\Z \]
for each choice of ${\lambda'}_\nu$.
 
\subsection{\it Summary}

In section \ref{SEC2} we will apply this construction to a few examples as
illustrations. Examples include group and oscillator algebras. In section
\ref{SEC3}, we realise the twisted algebra in terms of the elements of the
original algebra and certain ``charge'' operators $Q_\mu$. Such
realisations are known in the theory of quantum groups, for instance in
the $q$-oscillator realisation of $U_q(su(2))$.

Next in section \ref{SEC4} we construct two different deformed KM
algebras. First we deform the KM algebra using twisted oscillators.
Second we deform the KM generators directly. We then check that the KM
algebra remains ``Hopf'' after certain deformations as well, but with
twisted coproducts. At last we obtain the Virasoro algebra, via Sugawara
construction, using the second form of the deformed KM algebra. The
Virasoro algebra we obtain is the same as the usual one.

The deformation affects the $R$-matrix and the associated braid group
representations. That means that statistics is affected by twisting as was
emphasised in earlier papers \cite{FIORE,OECKL,GROSSE,ALBAL2, ALBAL3}. The ``abelian'' twist
based on $\R^N$ dealt with here does not change the permutation group
governing particle identity in the absence of twist. But it does change
the specific realisation of this group, with serious consequences for
physics. We discuss the twisted permutation symmetry in section
\ref{SEC5}.

We have acknowledged certain previous papers \cite{RIEFFEL,CONNESLANDI,CONNESDV}
for the origin of the ideas treated in this paper. It also has overlap
with the Fairlie-Zachos work on ``atavistic'' algebras \cite{ZACHOS}. Our
principal concern is the systematic construction of deformed algebras in
terms of undeformed ones, and that appears to originate from our own
previous work.  

The method used for deformation of the algebras presented in this papers
has similarities with a method used in works of Naihong Hu
\cite{HU1,HU2}. In particular, the deformed KM algebras we obtained in
the present work is the same as the ``color'' KM algebra obtained in \cite{HU2}.

\section{Examples of Abelian Twists}\label{SEC2}

As mentioned in the introduction, $(M,g)$ is a Riemannian manifold on
which $\R^N\;\;(N \geq 2)$ acts isometrically. The commutative algebra
$\calA(M)$ is the algebra of functions $\calC^\infty(M)$ with pointwise
multiplication. With scalar product induced by $g$, we can construct the
Hilbert space $L^2(M,d\mu_g)$ which can be decomposed as in
(\ref{eq:1}). $\calA(M) \equiv \calA_0(M) \subset L^2(M,d\mu_g)$
can then be twisted to $\calA_\theta(M)$ using the prescription
(\ref{eq:3}). We now give examples of such twisted algebras.

\subsection{\it The Moyal Plane $\calA_\theta(\R^{d+1})$}

In this case $\R^{d+1}$ acts on $\calA(\R^{d+1})=\calC^\infty(\R^{d+1})$
by translations leaving the flat Euclidean metric 
invariant. The
IRR's are labelled by the momentum $\lambda=(p^0,p^1,\ldots,p^d)$. A basis
for $\calH^{(p)}$ is formed by plane waves $e_p$ with $e_p(x) = e^{-ip_\mu
  x^\mu}$, $x = (x^0,x^1,\ldots,x^d)$ being a point of
$\R^{d+1}$. Following (\ref{eq:3}), the $*$-product is defined by
\begin{equation}
  \label{eq:4}
  e_p \ast_\theta e_q = e_p \; e_q \; e^{-\frac{i}{2}p_\mu \theta^{\mu \nu}q_\nu}\;.
\end{equation}
This $*$-product defines the Moyal plane $\calA_\theta(\R^{d+1})$.

\subsection{\it Functions on Tori}

This example is a compact version of the Moyal plane. The manifold $M$ is
the torus $T^N\;\;(N\geq 2)$. The group $\R^N$ acts via its homomorphic
image $U(1)\times U(1) \times \cdots \times U(1) \equiv U(1)^{\times N}$
on $T^N$ leaving its flat metric invariant. The IRR's are labelled by the
integral lattice $\Z^N$ with points $\lambda =
(\lambda_1,\ldots,\lambda_N),\;\;\lambda_i \in \Z$. A basis for
$\calH^{(\lambda)}$ is $e_\lambda$,
\[  e_\lambda(p) = e^{-i \lambda_i \theta^i}\;,\]
with $p = (e^{i\theta^1},\ldots,e^{i\theta^N})$ being a point of
$T^N$. The $*$-product is 
\begin{equation}
  \label{eq:5}
  e_\lambda *_\theta e_{\lambda'} = e_\lambda \; e_{\lambda'} \;
  e^{-\frac{i}{2} \lambda_\mu \theta^{\mu \nu} {\lambda'}_\nu}\;.
\end{equation}
It defines the noncommutative torus $T^N_\theta$.

\subsection{\it Functions on Groups}

Let $G=\set{g}$ be a simple, compact Lie group of rank $N \geq 2$ with
invariant measure $d\mu$ and let $T^N$ be its maximal torus. We can denote its
IRR's by  $\lambda =(\lambda_1,\ldots,\lambda_N),\;\;\lambda_i \in \Z$ as
before. $T^N$ can act on $G$ by left \emph{or} right multiplication. Let
us focus on the right action and the corresponding action on
$L^2(G,d\mu)$. As this action is unitary, we have the decomposition
(\ref{eq:1}). Now if $f_{\lambda},f_{\lambda'}\in
\calH^{(\lambda),(\lambda')}$ are two smooth functions
$f_{\lambda},f_{\lambda'}\in \calC^\infty(G)$, and $f_\lambda \otimes
f_{\lambda'} \rightarrow f_\lambda \; f_{\lambda'}$ is their pointwise
product, we can twist it:
\begin{equation}
  \label{eq:6}
  f_\lambda *_\theta f_{\lambda'} = f_\lambda \; f_{\lambda'} \;
  e^{-\frac{i}{2} \lambda_\mu \theta^{\mu \nu} {\lambda'}_\nu}\;.
\end{equation}

Now $T^N$ acts on the right \emph{and} left of $G$. That is, $T^N \times
T^N$ acts on $G$ and hence unitarily on $L^2(G,d\mu)$. We can use any of
its subgroups of rank $N \geq 2$ to perform the twist. With this
generalisation we can even choose $N=1$ and twist $\calA(S^3 \simeq
SU(2))$, which is $\calC^\infty(SU(2)) \simeq \calC^\infty(S^3)$ with
pointwise product.

We will be explicit about this twist. If $\set{D^J_{\lambda_1,\lambda_2}|
  J \in \set{0,\frac{1}{2},1,\ldots}}$ are the matrix elements of $SU(2)$
rotation matrices in the basis with third component of angular momentum
diagonal, we have the expansion
\begin{equation}
  \label{eq:7}
  f = \sum f^J_{\lambda_1,\lambda_2} D^J_{\lambda_1,\lambda_2}\;,
  \hspace{15mm} f \in \calC^\infty(G), \;\; f^J_{\lambda_1,\lambda_2} \in \C\;,
\end{equation}
by the Peter-Weyl theorem \cite{BALTRAH}. The twisted product is 
\begin{equation}
  \label{eq:8}
  D^J_{\lambda_1,\lambda_2}\;*\;D^K_{{\lambda'}_1,{\lambda'}_2} =
  D^J_{\lambda_1,\lambda_2}\;D^K_{{\lambda'}_1,{\lambda'}_2} \; e^{-\frac{i}{2}
    \lambda_\mu \theta^{\mu \nu} {\lambda'}_\nu}\;, \hspace{5mm}
  {\lambda}=({\lambda}_1,{\lambda}_2),\; {\lambda'}=({\lambda'}_1,{\lambda'}_2)\;.
\end{equation}

\subsection{\it Deforming Graded Algebras}\label{GRADALG}

Let $a_\lambda$ and $a_\lambda^\dagger$ $( \lambda =
(\lambda_1,\lambda_2,\ldots,\lambda_N), N \geq 2, \lambda_i \in \R)$ be
bosonic oscillators,
\[ \comm{a_\lambda}{a_{\lambda'}^\dagger} = \delta_{\lambda \lambda'},
\hspace{5mm} \comm{a_\lambda}{a_{\lambda'}} =
  \comm{a_\lambda^\dagger}{a_{\lambda'}^\dagger} = 0\;,\]
where $\delta$ is the Dirac delta and  $\delta_{\lambda \lambda'} =
\delta^{(N)}(\lambda - \lambda')$ if $\lambda_i, {\lambda'}_j$ take
continuous values. If we assign a charge $\lambda$ to $a_\lambda$ and
$-\lambda$ to $a_\lambda^\dagger$, then $a_\lambda a_{\lambda'}$ and 
$a_\lambda^\dagger a_{\lambda'}^\dagger$ have charges $\lambda + \lambda'$
and $-\lambda - \lambda'$ while $a_\lambda a_{\lambda'}^\dagger$ and
$a_{\lambda'}^\dagger a_\lambda$ have charges $\lambda - \lambda'$. Thus
$a_\lambda, a_{\lambda'}^\dagger$ generates a graded algebra $\calA$ with
charge $\lambda$ giving the grade:
\begin{equation}
  \label{eq:9}
  \calA = \bigoplus \calA^{(\lambda)}\;, \hspace{10mm}
  \calA^{(\lambda)}\calA^{(\lambda')} \subseteq \calA^{(\lambda + \lambda')}\;.
\end{equation}
This feature allows us to twist $\calA$ to an associative algebra
$\calA_\theta$ as before. Thus if $\alpha_\lambda \in \calA^{(\lambda)}$,
\begin{equation}
  \label{eq:10}
  \alpha_\lambda *_\theta \alpha_{\lambda'} =  \alpha_\lambda \;
  \alpha_{\lambda'} \;e^{\frac{i}{2} \lambda_\mu \theta^{\mu \nu} {\lambda'}_\nu} \;.
\end{equation}
In particular,
\begin{eqnarray}
a_\lambda *_\theta a_{\lambda'} =  a_\lambda \; a_{\lambda'}
\; e^{\frac{i}{2}\; \lambda_\mu \theta^{\mu \nu} {\lambda'}_\nu}\;,
&\hspace{10mm}& 
a_\lambda^\dagger *_\theta a_{\lambda'}^\dagger =  a_\lambda^\dagger \; a_{\lambda'}^\dagger
\; e^{\frac{i}{2}\; \lambda_\mu \theta^{\mu \nu} {\lambda'}_\nu}\;,
\nonumber \\  && \nonumber \\
a_\lambda *_\theta a_{\lambda'}^\dagger =  a_\lambda \; a_{\lambda'}^\dagger
\; e^{-\frac{i}{2}\; \lambda_\mu \theta^{\mu \nu} {\lambda'}_\nu}\;,
&\hspace{10mm}& 
a_\lambda^\dagger *_\theta a_{\lambda'} =  a_\lambda^\dagger \; a_{\lambda'}
\; e^{-\frac{i}{2}\; \lambda_\mu \theta^{\mu \nu} {\lambda'}_\nu}\;. \nonumber
\end{eqnarray}
Note that $\calA_\theta$ is graded:
\begin{equation*}
  \calA_\theta = \bigoplus \calA_\theta^{(\lambda)}\;.
\end{equation*}

Following the treatment of $\calA(S^3)$, we can twist even oscillators
with just one label $\lambda=\lambda_1$. In this case let $\lambda_1$ 
takes values $1,2$ as an example so that $a_i$ and
$a_i^\dagger$ are the Schwinger oscillators for $SU(2)$. Arrange them as a
matrix:
\begin{equation}
  \label{eq:11}
  \hat{g} = \left(
    \begin{array}{rr}
      a_1 & -a_2^\dagger \\
      a_2 & a_1^\dagger
    \end{array} \right) \;,
\end{equation}
$U(1) \times U(1)$ acts on $\hat{g}$:
\[ \hat{g} \rightarrow 
\left(
    \begin{array}{cc}
      e^{\frac{i\varphi}{2}} & 0 \\
      0  & e^{-\frac{i\varphi}{2}}
    \end{array} \right) \;\;\hat{g}\;\;
\left(
    \begin{array}{cc}
      e^{\frac{i\varphi}{2}} & 0 \\
      0  & e^{-\frac{i\varphi}{2}}
    \end{array} \right)\;.  \]
The $U(1)\times U(1)$ charges $q = (q_1,q_2)$ of $a_i,a_j^\dagger$ are 
\begin{center}
\begin{tabular}{c|cccc}
{} & $a_1$ & $a_2$ & $a_1^\dagger$ & $a_2^\dagger$ \\
\hline
$q=$ & $(1,1)$ & $(-1,1)$ & $(-1,-1)$ & $(1,-1)$ \\
\end{tabular}
\end{center}

So we can label $a_i, a_j^\dagger$ by $q$: \[ a_1 = A_{(1,1)}\;,
\hspace{5mm} a_2 = A_{(-1,1)}\;, \hspace{5mm} a_1^\dagger =
A^\dagger_{(-1,-1)}\;, \hspace{5mm} a_2^\dagger = A^\dagger_{(1,-1)}\;,
\]
the $q$-charge of $A^\dagger_q$ being $-q$.

The oscillators $A_q,A^\dagger_q$ are just like
$a_\lambda,a^\dagger_\lambda$ for $N\geq2$. Hence they can be twisted as previously.

\section{$\calA_\theta$ in Terms of $\calA_0$ and Charges}\label{SEC3}

In this section we show how to realise $\calA_\theta$ in terms of $\calA$
and certain charge operators. This construction was used 
fruitfully in previous papers \cite{ALBAL2} \cite{UVIR}.

The deformed product was previously given in terms of the undeformed
product. It is also possible to give it as a relation between elements of
$\calA_\theta$ if $\calA_0$ is commutative. Since 
\begin{eqnarray}
  a_\lambda *_\theta a_{\lambda'} &=&  a_\lambda \; a_{\lambda'}
\; e^{\frac{i}{2}\; \lambda_\mu \theta^{\mu \nu} {\lambda'}_\nu}\;, \nonumber \\
a_{\lambda'} *_\theta a_\lambda &=&  a_{\lambda'} \; a_{\lambda}
\; e^{-\frac{i}{2}\; \lambda_\mu \theta^{\mu \nu} {\lambda'}_\nu}\;, \nonumber
\end{eqnarray}
then for abelian {\bf $\calA_0$}, the relation is that of Weyl:
\[  a_\lambda *_\theta a_{\lambda'}  = e^{i\; \lambda_\mu \theta^{\mu
    \nu} {\lambda'}_\nu}\; a_{\lambda'} *_\theta a_\lambda \;.\]
Let us first discuss this simple case.

\subsection{\it Deformations of Abelian Algebras}\label{ABALG}

Let $Q_\mu$ be the charge operator:
\begin{equation}
  \label{eq:12}
  \comm{Q_\mu}{a_\lambda} = - \lambda_\mu \; a_\lambda\;.
\end{equation}
Set
\begin{equation}
  \label{eq:13}
  \hat{a}_\lambda = a_\lambda\;e^{-\frac{i}{2}\;\lambda_\mu \theta^{\mu
      \nu} Q_\nu}\;.
\end{equation}
Then under the $\theta=0$, unstarred products,
\begin{equation}
  \label{eq:14}
  \hat{a}_\lambda \; \hat{a}_{\lambda'} = a_\lambda \;
  a_{\lambda'}\;e^{-\frac{i}{2}\;(\lambda + \lambda')_\mu \theta^{\mu \nu}
    Q_\nu} e^{\frac{i}{2}\;\lambda_\mu \theta^{\mu \nu} {\lambda'}_\nu}\;.
\end{equation}
Hence
\begin{equation}
  \label{eq:15}
  \hat{a}_\lambda \; \hat{a}_{\lambda'} = e^{i\;\lambda_\mu \theta^{\mu
    \nu} \lambda_\nu}\; \hat{a}_{\lambda'}\;\hat{a}_\lambda
\end{equation}
so that $\hat{a}_\lambda$'s realise the $*_\theta$ algebra.

\subsection{\it The Case $\calA_0$ is Noncommutative}

We now argue that $\calA_\theta$ can always be realised as in \ref{ABALG},
also when $\calA_0$ is noncommutative.

Let $\calA_0 = \calA$ be a possibly noncommutative graded algebra as above: $\calA_0 = \oplus
\calA_0^{(\lambda)}$. To each $a_\lambda \in \calA_0^{(\lambda)}$, we
associate $\hat{a}_\lambda$ by the rule
\begin{equation}
  \label{eq:16}
  \hat{a}_\lambda = a_\lambda \; e^{-\frac{i}{2} \lambda_\mu \; \theta^{\mu
      \nu} \; Q_\nu}\;.
\end{equation}
Then
\begin{equation}
  \label{eq:17}
  \hat{a}_\lambda \; \hat{a}_{\lambda'} =
  \widehat{a_\lambda\;a_{\lambda'}}\;e^{\frac{i}{2} \lambda_\mu \;
    \theta^{\mu \nu} \; {\lambda'}_\nu}\;.
\end{equation}
Hence the image $\hat{\calA}_0$ of $\calA_0$ under the hat map is in fact
closed under multiplication, that is, is an algebra. It is also graded:
\begin{eqnarray}
  \hat{\calA}_0&=&\bigoplus_\lambda \hat{\calA}_0^{(\lambda)} \\
  \hat{\calA}_0^{(\lambda)}&=&\textrm{Image under hat of } \calA_0.
\end{eqnarray}
It is in fact a subalgebra of $\calA_\theta$ since we also have (\ref{eq:15})
from (\ref{eq:17}).

Conversely, if we define $a_\lambda$ by $a_\lambda=\hat{a}_\lambda
e^{\frac{i}{2}\lambda_\mu \theta^{\mu \nu} Q_\nu}$, then $a_\lambda$'s
fulfill the relations of $\calA_0$. The hat map being invertible, we
conclude that $\hat{\calA_0}=\calA_\theta$ and that the inverse map
$\hat{a}_\lambda\to a_\lambda$ gives $\calA_0$.

\section{Kac-Moody Algebras} \label{SEC4}

Now we discuss $SU(N)$ KM algebras. In the first instance, we assume that
$N \geq 3$, so that the rank of $SU(N)$ is at least $2$. The
(complexified) Lie algebra of $SU(N)$ has a basis $H_i, E_{\pm \alpha}$
where $H_i$ spans the Cartan subalgebra and $E_{\pm \alpha}$ are the
raising and lowering operators:
\[ \comm{H_i}{H_j} = 0 \;, \hspace{10mm} \comm{H_i}{E_{\alpha}} =
\alpha_i\;E_\alpha\;, \hspace{5mm}i,j \in \set{1,2,\ldots,N-1}\;. \] 

There is an oscillator construction of the KM algebra. We can twist the
oscillators which deforms the KM algebra. Or we can directly deform the KM
algebra. These
sets of deformations give different deformed KM algebras as we shall see.

\subsection{\it Oscillator Twists}\label{OSCTW}

Let $a_i,a_j^\dagger\;\;(1 \leq i,j \leq N)$ be bosonic annihilation and creation
operators. Then if $\lambda_a \;\;(a=1,2,\ldots,N^2 -1)$ are the $N \times
N$ Gell-Mann matrices of $SU(N)$, we have the Schwinger construction of
the $SU(N)$ Lie
algebra generators on the Fock space: 
\begin{eqnarray}
  \Lambda_\alpha &=& a^\dagger \;\lambda_\alpha \;a \nonumber \\
  \comm{\Lambda_a}{\Lambda_b} &=& i \; f_{abc}\;\Lambda_c\;.  \label{eq:18}
\end{eqnarray}
We can express $H_i,E_\alpha$ in terms of $\Lambda_a$ in a well-known way.

We can also label $a_j^\dagger$'s by weights $\mu^{(j)}$ such that
\begin{equation}
  \label{eq:19}
  \comm{H_i}{a^\dagger_{\mu^{(j)}}}= \mu_i^{(j)}\;a^\dagger_{\mu^{(j)}}\;.
\end{equation}
Then $a_j$ has weight $-\mu^{(j)}$. We can write it as $a_{\mu^{(j)}}$,
using the negative of the weights as subscript for $a$'s.

For the KM realisation we need infinitely many such oscillators pairs
${a^{(n)}_{\mu^{(j)}}}^\dagger,{a^{(n)}_{\mu^{(j)}}}\; (n=0,1,\ldots)$,  
${a^0_{\mu^{(j)}}}^\dagger, {a^0_{\mu^{(j)}}}$ being
$a^\dagger_{\mu^{(j)}}, a_{\mu^{(j)}}$. Their weights are $\pm \mu^{(j)}$:
\begin{eqnarray}
  \comm{H_i}{{a^{(n)}_{\mu^{(j)}}}^\dagger} &=& \mu^{(j)}_i \;
{a^{(n)}_{\mu^{(j)}}}^\dagger \nonumber \\ && \nonumber \\
  \comm{H_i}{{a^{(n)}_{\mu^{(j)}}}} &=& -\mu^{(j)}_i \;
{a^{(n)}_{\mu^{(j)}}}\;. \nonumber 
\end{eqnarray}
As $SU(N)$ now acts on all the oscillators, the new $\Lambda_a$ are
\begin{equation}
  \label{eq:20}
  \Lambda_a = \sum_{n \geq 0} a^{(n) \dagger} \; \lambda_a \; a^{(n)}\;.
\end{equation}

We can next do the twist:
\[ {a^{(n)}_{\mu^{(j)}}}^\dagger,{a^{(n)}_{\mu^{(j)}}} \longrightarrow 
{\hat{a}^{(n)\dagger}_{\mu^{(j)}}},{\hat{a}^{(n)}_{\mu^{(j)}}} \]
where
\begin{eqnarray}
\hat{a}^{(n)\dagger}_{\mu^{(j)}} &=& {a}^{(n)\dagger}_{\mu^{(j)}}
\;\;e^{\frac{i}{2}\mu_k^{(j)}\;\theta^{kl}\; H_l}\;, \label{eq:21} \\ && \nonumber \\
\hat{a}^{(n)}_{\mu^{(j)}} &=& {a}^{(n)}_{\mu^{(j)}}
\;\;e^{-\frac{i}{2}\mu_k^{(j)}\;\theta^{kl}\; H_l}\;. \label{eq:22}
\end{eqnarray}

Observe that $H_l$'s form a set of charge operators, i.e., $[H_i,H_j]=0$,
for all $i,j=1,...,N-1$. 

\subsection{\it The Kac-Moody Deformations}

As mentioned above, there are two ways to deform the KM algebra. The first
deformation is induced by those of the oscillators. In the second, we deform the
KM generators directly. They lead to different deformations of the KM algebra.

\subsubsection{From Twisted Oscillators}

The untwisted oscillators give the KM generators
\begin{eqnarray}
  J_a^{(n)} &=& \sum_m a^{(n+m)\dagger}\;\lambda_a \; a^{(m)} \equiv  \sum_m
  a^{(n+m)\dagger}_{\mu^{(j)}}\;(\lambda_a)_{jk} \; a^{(m)}_{\mu^{(k)}}\;,
  \label{eq:23} \\
  J _a^0 &\equiv& \Lambda_a\;, \label{eq:24} \\ 
a^{(r)}, a^{(r)^\dagger} &=& 0 \hspace{3mm}\textrm{if}\hspace{3mm} r<0\;. \nonumber 
\end{eqnarray}
We write the deformed KM generators $\hat{J}_a^{(n)}$ in terms of twisted oscillators
(\ref{eq:21}) and (\ref{eq:22}), so that:
\begin{equation}
  \label{eq:25}
  \hat{J}_a^{(n)} = \sum_m
  \hat{a}^{(n+m)\dagger}_{\mu^{(j)}}\;(\lambda_a)_{jk} \; \hat{a}^{(m)}_{\mu^{(k)}}\;.
\end{equation}
Substituting (\ref{eq:21}) and (\ref{eq:22}), we find
\begin{eqnarray}
  \hat{J}_a^{(n)} &=& \sum_m
  a^{(n+m)\dagger}_{\mu^{(j)}}\;(\lambda_a)_{jk}\;a^{(m)}_{\mu^{(k)}} \;\;  
e^{-\frac{i}{2}\mu_p^{(j)}\;\theta^{pq}\; \mu^{(k)}_q} \;\;e^{\frac{i}{2} (\mu^{(j)}-\mu^{(k)})_p\;\theta^{pq}\; H_q}
\nonumber \\
&=& \sum_m  a^{(n+m)\dagger}_{\mu^{(j)}}\; \left\{  (\lambda_a)_{jk}\;\;e^{-\frac{i}{2} \; \ad
    \overleftarrow{H}_p \; \theta^{pq} \; \overrightarrow{H}_q} \right\}  \;a^{(m)}_{\mu^{(k)}} 
\; e^{\frac{i}{2}(\mu^{(j)}-\mu^{(k)})_p\;\theta^{pq}\; H_q}\;, \label{eq:26}
\end{eqnarray}
where $\ad{H}_p\cdot = [H_p, \cdot ]$ is the adjoint action of $H_p$. 

The exponential in braces acts only on oscillators, in the manner already
shown. The Gell-Mann matrices are thus effectively changed to the
operators
\begin{equation}
  \label{eq:27}
  \hat{\lambda}_a = \lambda_a \; e^{-\frac{i}{2}\;\ad  \overleftarrow{H}_p \;
    \theta^{pq} \; \overrightarrow{H}_q} \;.
\end{equation}
We note that
\begin{equation}
  \label{eq:28}
  \comm{\lambda_a}{e^{-\frac{i}{2}\;\ad  \overleftarrow{H}_p \;
    \theta^{pq} \; \overrightarrow{H}_q}} = 0 \;.
\end{equation}
Hence
\begin{equation}
  \label{eq:29}
\comm{\hat{\lambda}_a}{\hat{\lambda}_b} = i \;{C_{ab}}^{c} \hat{\lambda}_c
\;e^{-i\;\ad  \overleftarrow{H}_p \; \theta^{pq} \; \ad \overrightarrow{H}_q}\;.
\end{equation}
Thus, we can write 
\begin{equation}
  \label{eq:30}
  \hat{J}_a^{(n)} = \sum_m
  {a}^{(n+m)\dagger}_{\mu^{(j)}}\;(\hat{\lambda}_a)_{jk} \;
  {a}^{(m)}_{\mu^{(k)}}\;e^{-\frac{i}{2}\;\ad  \overleftarrow{H}_p \; \theta^{pq} \;
    {H}_q}\;. 
\end{equation}

Now
\begin{eqnarray}
   \hat{J}_a^{(n)}\; \hat{J}_b^{(n')} &=& \sum_{\substack{m,m' \\
       j,j'\\ k,k'}}
   \left( {a}^{(n+m)\dagger}_{\mu^{(j)}}\;(\hat{\lambda}_a)_{jk} \;
   {a}^{(m)}_{\mu^{(k)}}\right)\;\left({a}^{(n'+m')\dagger}_{\mu^{(j')}}\;(\hat{\lambda}_b)_{j'k'} \;  
   {a}^{(m')}_{\mu^{(k')}}\right) \times \nonumber \\
&&  \hspace{15mm}\times \;e^{\frac{i}{2}\;(\mu^{(j)}-\mu^{(k)})_p \;
  \theta^{pq}\;(\mu^{(j')}-\mu^{(k')})_q }\;\times\;e^{-\frac{i}{2}\;\ad
  \overleftarrow{H}_p \; \theta^{pq} \;  {H}_q}\;,   \label{eq:31}
\end{eqnarray}
where $\ad \overleftarrow{H}_p$ is to be applied to all four oscillators, 
so that
\begin{eqnarray}
   \hat{J}_a^{(n)}\;\left( e^{\frac{i}{2}\;\ad
  \overleftarrow{H}_p \; \theta^{pq} \; \ad \overrightarrow{H}_q}\right)\;
\hat{J}_b^{(n')} &=&  \sum_{\substack{m,m' \\ j,j' \\k,k'}} \left( {a}^{(n+m)\dagger}_{\mu^{(j)}}\;(\hat{\lambda}_a)_{jk} \;
   {a}^{(m)}_{\mu^{(k)}}\right)\;\left({a}^{(n'+m')\dagger}_{\mu^{(j')}}\;(\hat{\lambda}_b)_{j'k'} \;  
   {a}^{(m')}_{\mu^{(k')}}\right)\;\times \nonumber \\
&& \hspace{30mm} \times \;\;e^{-\frac{i}{2}\;\ad
  \overleftarrow{H}_p \; \theta^{pq} \; {H}_q}\;.  \label{eq:32}
\end{eqnarray}
From this follows the deformed KM algebra:
\begin{eqnarray}
   \hat{J}_a^{(n)}\;\left( e^{i\;\ad
  \overleftarrow{H}_p \; \theta^{pq} \; \ad \overrightarrow{H}_q}\right)\;
\hat{J}_b^{(n')} &-&  \hat{J}_b^{(n')}\;\left( e^{i\;\ad
  \overleftarrow{H}_p \; \theta^{pq} \; \ad \overrightarrow{H}_q}\right)\;
\hat{J}_a^{(n)} = \nonumber \\ =
i\;{C_{ab}}^c \hat{J}_c^{(n+n')}\; &+& k\;n\;\delta_{ab}\;\delta^{n+n',0} \;e^{-i(\mu^{(j)} + \mu^{(j')})_p \;
  \theta^{pq} \; (\mu^{(k)} + \mu^{(k')})_q},  
  \label{eq:33}
\end{eqnarray}
with $k$, being the level of the KM algebra, is $1$ for the oscillator
construction. Observe that in this case the commutator defining the KM
algebra has been deformed.
\subsubsection{From Direct Deformation of KM Generators}

We can express the KM algebra in a new basis where $\lambda_a$ are
exchanged for $H_i,E_\alpha$ (in the defining representation). This then gives
KM generators $J_i^{(n)},J_\alpha^{(n)}$, $\alpha$ being $SU(N)$
roots. Roots being special instances of weights, we can also write the
basis as $J_i^{(n)},J_{\mu^{(s)}}^{(n)}$. Now we
dispense with oscillators and consider any level $k$ of the KM algebra. According to
our prescriptions, the deformed siblings of the KM basis elements are
\begin{eqnarray}
\tilde{J}_i^{(n)} &=& J_i^{(n)} \nonumber \\
\tilde{J}_{\mu^{(s)}}^{(n)} &=& J_{\mu^{(s)}}^{(n)} \;\; 
e^{\frac{i}{2}\; \mu_p^{(s)}\;\theta^{pq}\;H_q}\;. \label{eq:34.0}
\end{eqnarray}
If we put $E_{\mu^{(s)}}$ for $\lambda_a$, $\mu^{(s)}$  being a root,
then it has nonvanishing matrix elements only for $\mu^{(j)} - \mu^{(k)} =
\mu^{(s)}$. So (\ref{eq:34.0}) is almost the same as (\ref{eq:26}), but not quite. Equation
(\ref{eq:26}) has the extra phase 
\[ e^{\frac{i}{2}\; \mu_p^{(j)}\;\theta^{pq}\; \mu_q^{(k)}} \]
inside the sum. Substituting $\mu^{(j)} = \mu^{(k)}+\mu^{(s)}$ and using
the antisymmetry of $\theta^{pq}$, this simplifies to 
\[ e^{\frac{i}{2}\; \mu_p^{(s)}\;\theta^{pq}\; \mu_q^{(k)}}\;. \]
So it appears that the two deformations are different.

The algebraic structure of $\tilde{J}_{\mu^{(s)}}^{(n)}$ is
simple. Clearly
\begin{eqnarray}
  \comm{\tilde{J}_i^{(n)}}{\tilde{J}_j^{(m)}} &=&
  k\;n\;\delta_{ij}\;\delta^{n+m,0} \label{eq:34} \\
\comm{\tilde{J}_i^{(n)}}{\tilde{J}_{\mu^{(s)}}^{(m)}},
&=& \mu^{(s)}_i \;\; \tilde{J}_{\mu^{(s)}}^{(n+m)}\;. \label{eq:35}
\end{eqnarray}
Also
\begin{equation}
  \label{eq:36}
  \comm{\tilde{J}_{\mu^{(s)}}^{(n)}\;e^{-\frac{i}{2}\;
      \mu_p^{(s)}\;\theta^{pq}\;H_q}}{\tilde{J}_{\mu^{(t)}}^{(m)}\;e^{-\frac{i}{2}\; \mu_{p'}^{(t)}\;\theta^{p'q'}\;H_{q'}}} = \comm{J_{\mu^{(s)}}^{(n)}}{J_{\mu^{(t)}}^{(m)}} = N_{st}\;\;
J_{\mu^{(s)}+\mu^{(t)}}^{(n+m)}
\end{equation}
where
\[ N_{st} \neq 0 \Longleftrightarrow  \mu^{(s)}+\mu^{(t)} \;\;\textrm{is a
  root}\;. \]
Hence, the LHS of (\ref{eq:36}) is equal to
\begin{equation}
 N_{st} \;\; \tilde{J}_{\mu^{(s)}+\mu^{(t)}}^{(n+m)} \;\;
e^{-\frac{i}{2}\left( \mu^{(s)}+\mu^{(t)}\right)_p \;\theta^{pq}\;H_q}\;.  \label{eq:35.1}
\end{equation}

\subsubsection{Hopf Structure of Deformations}

The existence of a coproduct $\Delta$ is an essential property of a
symmetry algebra. With the help of $\Delta$, we can compose subsystems
transforming by the symmetry algebra and define the action of the latter
on the composite system. An algebra with a coproduct $\Delta$ is known as
a coalgebra.

If the symmetry algebra has a more refined structure and is Hopf (and not
just a coalgebra), then it has all the essential features of a group. In
this case, we can regard it as a ``quantum group of symmetries''
generalising ``classical'' symmetry groups \cite{MACKSCHOM}.

General deformations of a Hopf algebra such as the KM algebra need not preserve its Hopf
structure. We now show that $\tilde{J}_{\mu^{(s)}}^{(m)}$ is
in fact a basis of generators for a Hopf algebra. The situation as regards
$\hat{J}_a^{(n)}$ is less clear.

Both $J_{\mu^{(s)}}^{(m)}$ and $H_q$ are elements of a Hopf algebra. In
fact $J_{\mu^{(s)}}^{(0)}$ and $H_q$ generate (complexified) $su(N)$ Lie
algebra. From the expression for $\tilde{J}_{\mu^{(s)}}^{(m)}$, we see that
$\tilde{J}_{\mu^{(s)}}^{(m)}$ is also an element of the same
Hopf algebra establishing the claim. If $\Delta$ is the coproduct, we can
write
\begin{equation}
  \label{eq:37}
  \Delta\left(\tilde{J}_{\mu^{(s)}}^{(m)}\right) =
  \Delta\left(J_{\mu^{(s)}}^{(m)}\right) \; e^{\frac{i}{2}\;\mu^{(s)}_p \;
      \theta^{pq} \; \Delta(H_q)}\;, 
\end{equation}
$\Delta$ on KM generators having familiar expressions such as
\[ \Delta(H_q) = \mathbf{1} \otimes H_q + H_q \otimes \mathbf{1}\;. \]

As regards $\hat{J}_a^{(n)}$, we do not have an answer. They do not
seem to be elements of the enveloping algebra of the KM algebra.

\subsubsection{Remarks}

In both the oscillator and KM deformations, there is a superscript such as $(n)$
identifying the mode. It is passive in the process of deformation: the
antisymmetric deformation matrix $\theta=\left(\theta^{\mu \nu}\right)$ is independent of
$n$.

We can make it depend on $n$. We can replace $\theta$ by
$\theta^{(n)}=\theta^{(n)}_{\mu \nu}$ in the preceding construction,
thereby obtaining very general deformations. We will not study this
generalisation in this paper.

We can also deform $SU(2)$ KM and its Virasoro algebras by twisting the
Schwinger oscillators of $SU(2)$ following section \ref{GRADALG}. That
leads to the deformed $SU(2)$ KM currents similar to
$\hat{J}^{(m)}_{\mu^{(s)}}$ above.

\subsection{\it Virasoro Algebra}
 
The Virasoro algebra can be realised from the KM algebra by the Sugawara
construction. Its generators $L_n$ can be written as 
\begin{equation}\label{eq:37.1}
 L_n = \frac{1}{2k+C_N} : J_a^{(m+n)} \; J_a^{(-m)} : \;,
\end{equation}
where $k$ is the level of the KM algebra, and $C_N$ is the eigenvalue of the Casimir
operator of
$SU(N)$.

Here $:\hspace{3mm}:$ denotes normal ordering with regards to the
currents $J_a$. Those with positive superscripts stand to the right, and
we have reverted to the original subscripts $a$. Deformed currents in RHS
of (\ref{eq:37.1}) will then deform the Virasoro algebra as well.

The central role of the Virasoro algebra in physics is as a symmetry
algebra. That suggests that its deformation from $\tilde{J}_a^{(n)}$
is more interesting. So we focus on the deformation from
$\tilde{J}_{\mu^{(s)}}^{(n)}$. They deform $L_n$ to
\begin{equation}
\tilde{L}_n = \frac{1}{2k+C_N} \sum_m: \tilde{J}_a^{(m+n)} \;
\tilde{J}_a^{(-m)} : \nonumber
\end{equation}
Using (\ref{eq:34.0}), we can write
\begin{equation}
  \label{eq:38}
  \tilde{L}_n  =  \frac{1}{2k+C_N} \sum_m: J^{(m+n)}_{\mu^{(s)}}\;
J^{(-m)}_{-\mu^{(s)}} + J^{(m+n)}_{i}\; J^{(-m)}_{i}:\;,
\end{equation}
where we used $\mu^{(s)}_p~\theta^{pq}~(-\mu^{(s)}_q)=0$. Thus in this approach the Virasoro algebra is not deformed at all. 

For an implementation of the quantum conformal invariance in the $2-d$
Moyal plane see \cite{LIZZI}.

\section{On Statistics}\label{SEC5}

Suppose we have a free quantum scalar field $\varphi$ on the commutative
manifold $\R^{d+1}$ with the Fourier expansion
\[ \varphi(x) = \int d\mu(p) \left[ c(p)\;e_p(x) + c^\dagger(p)\;e_{-p}(x)
\right]\;, \] 
\[ p\cdot x = p_0\;x_0 - \vec{p}\cdot\vec{x}\;, \hspace{5mm}
e_p(x)=e^{-i\;p\cdot x}\;,\hspace{5mm} d\mu(p) =\frac{1}{(2\pi)^{\frac{d}{2}}}
\frac{d^d p}{2|p_0|}\;, \hspace{5mm} p_0=\sqrt{\vec{p}^2 + m^2}\;, \] \label{eq:stat}
where $x_0$ and $\vec{x}$ are the time and space coordinates, and 
$m$ is the
mass of $\varphi$. The creation and annihilation operators fulfill the
standard commutation relations
\begin{eqnarray*}
 \comm{c(p)}{c^\dagger(p')} &=& 2 \; |p_0| \; \delta^d (p'-p), \\
 \comm{c^\dagger(p)}{c^\dagger(p')} &=& \comm{c(p)}{c(p')} = 0.  
\end{eqnarray*}

We can now twist $c(p)$ and $c^\dagger(p)$ to
\begin{equation}
a(p) =c(p)\;e^{-\frac{i}{2} \; p_\mu \;\theta^{\mu \nu}\;P_\nu}\;, \hspace{10mm}
a^\dagger(p) = c^\dagger(p)\;e^{\frac{i}{2} \; p_\mu \;\theta^{\mu
    \nu}\;P_\nu} 
\end{equation} 
where $p_0$ and $\vec{p}$ are energy and momentum and $P_\mu$ is the
translation operator,

\begin{eqnarray} 
P_\mu:=\int d\mu(p) p_\mu c^\dagger(p)c(p)&=&\int d\mu(p) p_\mu
a^\dagger(p)a(p), \\
\comm{P_\mu}{a^\dagger(p)}=p_\mu\;a^\dagger(p)\;,& &
\comm{P_\mu}{a(p)} = -p_\mu\;a(p)\;.
\end{eqnarray}
The $P_\mu$ operator is the analogue of $Q_\mu$. It was studied in
\cite{OURPAPERS,ALBAL2,UVIR,ALBAL3}. The twist of $c$'s twists statistics since $a$'s and
$a^\dagger$'s no longer fulfill standard relations:
\[ a(p)\;a(p') = a(p')\;a(p)\;e^{i\;p_\mu \; \theta^{\mu\nu}
  p'_\nu}\;, \hspace{10mm}
   a^\dagger(p)\;a^\dagger(p') = a^\dagger(p')\;a^\dagger(p)\;\;e^{i\;p_\mu \; \theta^{\mu\nu}
  p'_\nu}\;, \]
\[   a(p)\;a^\dagger(p') = 2|p_0|\;\delta^d(p - p') +
   a^\dagger(p')\;a(p)\;e^{-i\;p_\mu\;\theta^{\mu \nu}\;p'_\nu}\;.\]

The implication of this twist is that the $n$-particle wave function
$\psi_{k_1 \cdots k_n}$,
\[ \psi_{k_1,\cdots,k_n}(x_1,\ldots,x_n) = \bra{0}\varphi(x_1)\varphi(x_2)\ldots \varphi(x_n)~a^\dagger_{k_n}a^\dagger_{k_{n-1}}\ldots
a^\dagger_{k_1}\;\ket{0} \]
is not symmetric under the interchange of $k_i$. Rather it fullfils a
twisted symmetry:
\begin{equation}\label{eq:42.0}
 \psi_{k_1 \cdots k_i\;k_{i+1} \cdots k_n} = e^{-i\; k_{i,\mu}
  \; \theta^{\mu \nu} \; k_{i+1,\nu}} \;  \psi_{k_1 \cdots k_{i+1}\;k_i
  \cdots k_n}\;. 
\end{equation}
This twisted statistics by the following chain of connections implies that
spacetime is the Moyal plane with
\begin{equation}
  \label{eq:42}
e_p \ast_\theta e_{p'} = e^{-\frac{i}{2}\;p_\mu\;\theta^{\mu \nu}\;p'_\nu}\;\;e_{p+p'}\;.   
\end{equation}

The chain is as follows: Let $g$ be an element of the Lorentz group
without time-reversal. For $\theta^{\mu \nu} = 0$, it acts on $\psi_{k_1 \cdots k_n}$ by
the representative $g\otimes g \otimes \cdots \otimes g$ ($n$ factors)
compatibly with the symmetry of $\psi_{k_1 \cdots k_n}$. This action is
based on the coproduct
\[ \Delta_0(g) = g \otimes g\;.\] 

But for $\theta^{\mu \nu} \neq 0$, and for $g\neq\textrm{identity}$, already for $n=2$,
\begin{eqnarray*}
 \Delta_0(g)\;\psi_{k_1,k_2} &=&\;\psi_{gk_1,gk_2}\;= e^{-i\;k_{1,\mu} \; \theta^{\mu
    \nu} \; k_{2,\nu}} \; \Delta_0(g) \;\psi_{k_2,k_1}\;=\\ 
                            &=& e^{-i\;k_{1,\mu} \; \theta^{\mu
    \nu} \; k_{2,\nu}} \;\psi_{gk_2,gk_1}\;\neq e^{-i\;(g~k_1)_\mu \; \theta^{\mu
    \nu} \; (g~k_2)_\nu} \;\psi_{gk_2,gk_1}\;.
\end{eqnarray*}

Thus the naive coproduct $\Delta_0$ is incompatible with the statistics
(\ref{eq:42.0}). It has to be twisted to
\begin{equation}
  \label{eq:43}
  \Delta_\theta(g) = F^{-1}_\theta(g\otimes g)F_\theta\;, \hspace{5mm}
  F_\theta = e^{\frac{i}{2}\partial_\mu \otimes \theta^{\mu \nu} \partial_\nu}\hspace{5mm} 
\end{equation}
for such compatibility.

But then $\Delta_\theta$ is incompatible with the commutative
multiplication map $m_0$:
\[ m_0(e_p \otimes e_{p'}) = e_{p+p'}\;. \]
That is, 
\[ m_0 \left[ \Delta_\theta(g)(e_p\otimes e_{p'}) \right] \neq g \;\;
e_{p+p'}\;. \]
We are forced to change $m_0$ to 
\[ m_\theta = m_0 \; F_\theta \]
for this compatibility, that is, to preserve spacetime symmetries as
automorphisms. Since
\begin{equation}
  \label{eq:44}
  m_\theta(e_p \otimes e_{p'}) = e_p \ast e_{p'} \equiv e^{-\frac{i}{2}\;
    p_\mu \; \theta^{\mu \nu} \; p'_\nu}\; e_{p+p'} \;, 
\end{equation}
we end up with the Moyal plane.

Thus statistics can lead to spacetime noncommutativity. This idea is being studied
further by our group.

In general, when we twist creation and annihilation operators such as $a_\lambda^\dagger$
and $a_\lambda$, then we twist statistics just as in the Moyal case
(\ref{eq:42}). The spatial slice associated with these operators can be the
$N$-torus $T^N$ if $\lambda_i \in \Z$, with coordinates
$(e^{i\;\theta_1},e^{i\;\theta_2},\ldots,e^{i\;\theta_N})$. The field
operator at a fixed time is then $\varphi$ where
\begin{equation}
  \label{eq:45}
  \varphi(e^{i\;\theta_1},e^{i\;\theta_2},\ldots,e^{i\;\theta_N}) =
  \sum_{\{\lambda_1\}} \left[ a_\lambda \; e^{-i \sum_i \lambda_i \theta_i}
    + a^\dagger_\lambda \; e^{i \sum_i \lambda_i \theta_i} \right]\;,
\end{equation}
where we have assumed for simplicity that $\varphi^\dagger=\varphi$.
Then the torus algebra is twisted for the same reason as in the Moyal
case. If $e_\lambda$ denotes the function with 
values
\[ e_\lambda(e^{i\;\theta_1},e^{i\;\theta_2},\ldots,e^{i\;\theta_N}) =  e^{-i
  \sum_i \lambda_i \theta_i}\;, \]
then their product $\ast$ is defined by
\begin{equation}
  \label{eq:46}
e_\lambda \ast e_{\lambda'} = e^{i\;\lambda_\mu \; \theta^{\mu \nu}\;
  \lambda'_\nu}\;e_{\lambda + \lambda'}\;.  
\end{equation}
That is we get back the twisted $\calC^\infty(T^N)$ algebra of
(\ref{eq:5}).

If there is a collection of oscillators indexed by $n$ as in section \ref{OSCTW}, or
equally KM generators with index $n$, it is more  reasonable to regard
them as associated $\calC^\infty(S^1)$. For example,
\begin{eqnarray}
  J_a(\theta) &=& \sum J_a^{(n)} \; e_n(e^{i\;\theta}) \;, \nonumber \\
  e_n(e^{i\;\theta}) &=& e^{-i\;n\;\theta}\;. \nonumber
\end{eqnarray}
This expansion is the known one for the generators of the Lie algebra of
the centrally extended loop group.

In this case, $a$ becomes an internal index. There is perhaps still an
interpretation of the deformation in terms of statistics of ``internal''
excitations associated with $a$. But $\calC^\infty(S^1) =
\calC^\infty(T^1)$ cannot be deformed like $\calC^\infty(T^N)$ for $N \geq
2$. So what these deformations have to do with spacetime twists is not clear.

\section{Final Remarks}

As remarked earlier, deformations like those we consider appeared first in
quantum group theory. Recently, they found concrete applications in
discussions of quantum theories on the Moyal plane and in particular Pauli
principle violations and the absence of UV-IR mixing
\cite{ALBAL2,UVIR,ALBAL3}. Further applications exist. The twists of the Moyal plane
are those of the world volume. We can also twist the target of fields with
striking results. Work on such twists is now being written up \cite{ALBAL}.

As mentioned earlier, recently, Fairlie and Zachos proposed an ``atavistic''
algebra \cite{ZACHOS}, which is based on the oscillator algebra. They also
called attention to the possible quantum field theoretical applications of
their algebra.
 
\section{Acknowledgements}

We thank Prof. V. Dobrev for comments on nomenclature of Hopf algebras. We
also thank Prof. David B. Fairlie for calling our attention to the work on
atavistic algebras. We also thank Prof. Naihong Hu for calling our
attention to his work and for some interesting comments on our work. The
work of APB is supported in part by DOE under grant number
DE-FG02-85ER40231. ARQ was supported by FAPESP under grant number
02/03247-2.



\begin{thebibliography}{10}

\bibitem{RIEFFEL} Rieffel M.A., ``Deformation Quantization for Actions
    of $\R^d$'', \emph{Memoirs of American Mathematical Society},
    \textbf{506} (AMS, Providence, 1993).
%
\bibitem{CONNESLANDI} Connes A., Landi G. ``Noncommutative Manifolds, the
    Instanton Algebra and Isospectral Deformations'',
    \emph{Comm. Math. Phys.} \textbf{221}, 1 (1993) 141.
%
\bibitem{CONNESDV} Connes A., Dubois-Violette M.,
``Noncommutative Finite-Dimensional Manifolds. I. Spherical Manifolds and
Related Examples'', \emph{Comm. Math. Phys.} \textbf{230}, 3 (2002) 539.  
%
%
\bibitem{FIORE} Fiore G., Schupp P., ``Identical Particles and
  Quantum Symmetries'', \emph{Nuc. Phys.} \textbf{B470}, 1-2 (1996)
  211; arXiv: hep-th/9508047. 
%
\bibitem{OECKL} Oeckl R., ``Braided Quantum Field Theory'',
  \emph{Comm. Math. Phys.} \textbf{217}, 2 (2001) 451; arXiv: hep-th/9906225.  
%
\bibitem{GROSSE} Grosse H., Maceda
  M., Madore J., Steinacker H., ``Fuzzy Instanton'',
  \emph{Int. J. Mod. Phys.} \textbf{A17} (2002) 2095; arXiv: hep-th/0107068.
%
\bibitem{ALBAL2} A.P. Balachandran, G. Mangano, A. Pinzul, S. Vaidya,
  ``Spin and Statistics on the Groenewold-Moyal Plane: Pauli-Forbidden
  Levels and Transitions'', Int.J.Mod.Phys. A21 (2006) 3111-3126; arXiv: hep-th/0508002.
%
\bibitem{ALBAL3} A. P. Balachandran, T. R. Govindarajan, G. Mangano,
  A. Pinzul, B. A. Qureshi, S. Vaidya, ``Statistics and UV-IR Mixing with
  Twisted Poincare Invariance'', Phys. Rev. D (in press); arXiv: hep-th/0608179.
%
\bibitem{ZACHOS} Fairlie, D. B. and Zachos, C. K., ``An Atavistic Lie
  Algebra'' arXiv: hep-th/0603017.

%
\bibitem{HU1} Hu, N., ``Quantum divided power algebra, q-derivatives and some new
quantum groups'', Journal of Algebra 232(2) (2000),507--540.

%
\bibitem{HU2} Hu, N., ``Quantum group structure associated to the quantum affine
space'', Algebra Colloquium 11(4) (2004), 483--492.
%
\bibitem{BALTRAH} Balachandran A.P., Trahern C.G., ``Lectures on Group
  Theory for Physicists'', (Bibliopolis, Napoli, 1984).
%
%
\bibitem{UVIR} Balachandran A.P., Pinzul A., Qureshi B., ``UV-IR Mixing in
  Non-Commutative Plane'', Phys.Lett. B634 (2006) 434-436; arXiv: hep-th/0508151.
%
\bibitem{MACKSCHOM} Mack G., Schomerus V., ``Quasi Hopf Quantum Symmetry
  in Quantum-Theory'', \emph{Nucl. Phys.} \textbf{B370}, 1 (1992) 185. 
%
%
\bibitem{LIZZI} Lizzi, F., Vaidya, S., Vitale, P., ``Twisted Conformal
  Symmetry in Noncommutative Two-Dimensional Quantum Field Theory''
  Phys.Rev. D73 (2006) 125020; arXiv: hep-th/0601056. 
%
\bibitem{OURPAPERS} Balachandran A.P., Govindarajan T. R., Harikumar E.,
  Queiroz A.R., Teotonio-Sobrinho P., (in preparation)
%
%
\bibitem{ALBAL} Balachandran A.P., Marques A.M., Teotonio-Sobrinho P.,
 ``Deforming an Abelian Scalar Theory'' (in preparation). 
%

\end{thebibliography}
\end{document}